\documentclass[11pt]{article}
\usepackage[margin=1in]{geometry}
\usepackage{amsmath, amssymb}
\usepackage{graphicx}
\usepackage{cite}        
\usepackage{authblk}     
\usepackage{hyperref}    

\usepackage{graphicx}
\usepackage{dcolumn}
\usepackage{mathtools}
\usepackage{bm}
\usepackage{hyperref}
\usepackage{xcolor}
\usepackage[caption=false]{subfig}

\usepackage{multirow}
\usepackage{cleveref}
\usepackage{rotating}
\usepackage{adjustbox}
\usepackage{tikz}
\usetikzlibrary{shapes}
\usepackage{amsmath}

\usepackage{etoolbox}
\usetikzlibrary{plotmarks}
\usepackage{MnSymbol}
\DeclareRobustCommand\full  {\tikz[baseline=-0.6ex]\draw[thick] (0,0)--(0.5,0);}
\DeclareRobustCommand\dotted{\tikz[baseline=-0.6ex]\draw[thick,dotted] (0,0)--(0.54,0);}
\DeclareRobustCommand\dashed{\tikz[baseline=-0.6ex]\draw[thick,dashed] (0,0)--(0.54,0);}

\DeclareRobustCommand\dashdot {\tikz[baseline=-0.6ex]\draw[thick,dash dot] (0,0)--(0.5,0);}

\definecolor{darkgreen}{RGB}{0.0, 100, 0.0}
\definecolor{blueplot}{RGB}{12.0, 93, 165}
\definecolor{greenplot}{RGB}{0.0, 185, 69}
\definecolor{orangeplot}{RGB}{255, 149, 0.0}
\definecolor{brownplot}{RGB}{140, 86, 75}
\definecolor{grey}{RGB}{127, 127, 127}

\newcommand\marksymbol[3]{\tikz[draw = #2, fill = white, scale = #3]\pgfuseplotmark{#1};}
\newcommand\diams[2]{\tikz[draw = #1, scale = #2]\pgfuseplotmark{diamond*};}

\newcommand\myeq{\mathrel{\underset{\makebox[0pt]{\mbox{\normalfont\tiny\sffamily $\tau \rightarrow \infty$}}}{=}}}

\DeclareMathOperator{\Rr}{1/\mathcal{E}}
\DeclareMathOperator{\Rrstar}{1/\mathcal{E}_*}

\newcommand\norm[1]{\left\lVert#1\right\rVert_2}
\newcommand*\diff{\mathop{}\!\mathrm{d}}
\newcommand{\avg}[1]{\left<#1\right>}

\title{\bfseries Numerical investigations around the Gallavotti-Cohen Fluctuation Theorem on Log-lattices}

\author{G. Costa\textsuperscript{1,2} and Bérengère Dubrulle\textsuperscript{1\dag}}

\date{}

\begin{document}
\maketitle
\vspace*{-2em}
\noindent\makebox[0pt][r]{%
  \begin{tikzpicture}[remember picture, overlay]
    \node[anchor=north east, xshift=-1cm, yshift=-0.5cm] at (current page.north east) {%
      \footnotesize\textbf{Submitted to \textit{Philos. Trans. R. Soc. A}}};
  \end{tikzpicture}%
}
\vspace{-1.5em}  
\begin{center}
    \textsuperscript{1}
    Universite Paris-Saclay, CEA, CNRS, SPEC, 91191, Gif-sur-Yvette, France\\ 
    \textsuperscript{2}Université Côte d’Azur, Observatoire de la Côte d’Azur, CNRS, Laboratoire Lagrange, France\\
    \vspace{0.5em}
    \textsuperscript{\dag} \textit{Corresponding author:} \texttt{berengere.dubrulle@cea.fr}
\end{center}

\vspace{-1em}

\maketitle

\begin{abstract}
Using the recent concept of fluids projected onto \textit{Log-Lattices}, we investigate the validity of the Gallavotti-Cohen Fluctuation Theorem (GCFT) in the context of fluid mechanics. The dynamics of viscous flows are inherently irreversible, which violates a fundamental assumption of the fluctuation theorem. To address this issue, Gallavotti introduced a new model, \textit{the Reversible Navier-Stokes Equation} (RNS), which recovers the time-reversal symmetry of the Navier-Stokes (NS) equations while retaining the core characteristics of the latter. We show that \textit{for fluids on Log-Lattices}, the GCFT holds for the RNS system. Furthermore, we show that this result can be extended, under certain assumptions, to the traditional, irreversible Navier-Stokes equations. Additionally, we show that the phase space contraction  rate satisfies a large deviation relation which rate function can be estimated.
\smallbreak\noindent\textbf{Keywords:} Fluctuation theorem, Navier-Stokes, Reversibility
\end{abstract}

\section{Introduction}
While macroscopic dynamics generally emerge from reversible microscopic descriptions of $N$-particle systems, they are still subject to the well-known Second Law of Thermodynamics, insofar as they are governed by irreversible equations that produce entropy, which is, on average, positive. In the context of dynamical systems, this positive mean entropy production is associated with a positive mean contraction rate of the phase space, and the possibility of development of an attractor. A first characterization of the statistics of entropy production was provided by Sinai~\cite{sinai1977lectures}, who showed that $Y$, the mean entropy production over a period $\tau$ obeys a large deviation principle:
\begin{equation}
        \label{eq:LargeDevInf}
        P_\tau(Y = y) = A_\tau e^{\zeta(y)\tau},
\end{equation}
\noindent where $\zeta$ is the large deviation function.  

\medbreak\noindent A second characterization of the statistics of entropy production was provided in 1993 by Evans et al.~\cite{evans1993probability} for dynamical systems that are reversible (in the sense of \textit{Anosov systems}). In such cases, signatures of reversibility may persist statistically, leading to dynamical fluctuations that violate the Second Law of Thermodynamics. The probability of such events is quantified through an \textit{universal} relation (i.e., without free parameters) that connects the stationary state probability of observing a given value of the average entropy production rate to the probability of observing its opposite. This relation is now known as the Fluctuation Theorem (FT), and it can be used to constrain the shape of the large deviation function.

\medbreak\noindent Anosov systems, however, are highly idealized and do not represent most physical systems encountered in nature, which are typically characterized by irreversible fluxes (e.g., heat, energy, etc.). As a first step toward generalizing to physical systems, Gallavotti and Cohen~\cite{gallavotti1995dynamical} proposed in 1995 an alternative version of the Fluctuation Theorem (FT), based on a \textit{Chaotic Hypothesis} (CH). This hypothesis posits that strongly chaotic systems can be treated as Anosov systems. Since this hypothesis originates from turbulence (and was similarly formulated by Ruelle in~\cite{ruelle1995measures}), turbulent fluids appear as ideal candidates for the study of the Gallavotti-Cohen Fluctuation Theorem (GCFT). This motivated several investigations of the GCFT using data from turbulent flows~\cite{ciliberto1998experimental,ciliberto2004experimental}.
The main challenge in the direct application of the theorem lies in the fact that viscous fluids break the time-reversal symmetry at the macroscopic level (for the Navier-Stokes equations), meaning that entropy production (via viscous dissipation) is always positive, thus prohibiting a straightforward application of the FT.

\medbreak\noindent This problem was circumvented in~\cite{ciliberto1998experimental,ciliberto2004experimental} by using the "average of local observables",  based on a conjecture by Gallavotti~\cite{gallavotti1996equivalence} stating that these observables could also follow the GCFT. Specifically, it was shown that the time-averaged heat transfer in Rayleigh-Bénard convection~\cite{ciliberto1998experimental}, the integral of the pressure over a small plate in a swirling closed flow, and the force applied to a small plate inside a wind tunnel~\cite{ciliberto2004experimental}, all followed a generalization of the FT. In~\cite{ciliberto2004experimental}, probability density functions (PDFs) were also successfully rescaled, applying Sinai's theory~\cite{sinai1977lectures}, leading to $\tau$ -independent distributions. However, since these systems are far from the initial assumptions of the GCFT, it remains unclear whether this is merely a coincidence or if there is a deeper connection with the breaking of time-reversibility.

\medbreak\noindent In this paper, we propose an alternative approach to circumvent the problem by using the Reversible Navier-Stokes (RNS) equations. In this framework, viscosity is no longer constant but fluctuates in time to conserve the total enstrophy~\cite{gallavotti1995dynamical, margazoglou2022nonequilibrium} or the total kinetic energy~\cite{shukla2019phase, costa2023reversible, CostaRNS24}. The regained time-reversibility, combined with the Chaotic Hypothesis (CH), allows for the controlled use of the Gallavotti-Cohen Fluctuation Theorem (GCFT), which could provide new insights into the Navier-Stokes equations. Indeed, the RNS equation not only restores time-reversibility but also satisfies~\cite{margazoglou2022nonequilibrium, costa2023reversible} the equivalence conjecture~\cite{gallavotti1996equivalence}, which states the statistical equivalence between the reversible RNS and the irreversible Navier-Stokes (NS) system.
However, the application of the GCFT in this context is still impeded by the requirement to reach very long time scales in order to probe rare fluctuations. This limitation makes it difficult to explore the problem using direct numerical simulations, as the computational cost scales as $\nu^{-3}$ and becomes prohibitively large for fluctuations relevant to the GCFT, which correspond to $\nu\to 0$ (or even negative). To make the problem tractable, we therefore resort to the projection of the fluid equations onto log-lattices (see Appendix~\ref{Sec:LL}), a method that enables the simulation of flows at very large Reynolds numbers at an affordable computational cost~\cite{campolina2021fluid}.

\section{Background}
     In this section, we provide some background material regarding the GCFT, large deviations and the Reversible Navier-Stokes system.
    \subsection{The Gallavotti-Cohen Fluctuation Theorem}
            \label{sec:GCFT_sec}

            The GCFT applies to reversible many-particle systems in a stationary state, provided they follow the \textit{Chaotic Hypothesis}~\cite{gallavotti1995dynamical}, which is less restrictive than the Anosov condition. Schematically, the Chaotic Hypothesis assumes that a system exhibiting chaotic motion does so in a maximal form so that it can be considered a transitive hyperbolic system. We can then state the following theorem:
            
            \medbreak {\bf CHA}: \textit{A chaotic, reversible many-particle system in a stationary state can be regarded as a transitive Anosov system for the purpose of computing the macroscopic properties of the system.}
            
            \medbreak\noindent Through the use of (CHA), it becomes possible to apply the results from Evans et al.~\cite{evans1993probability} and derive the \textit{Gallavotti-Cohen Fluctuation Theorem} (GCFT).

            \medbreak\textbf{Gallavotti-Cohen Fluctuation Theorem (GCFT):}
            \medbreak\noindent If a system satisfies the following properties (A)-(C):
            
            \medbreak (A) \textit{Dissipation:} The phase space volume undergoes a contraction at an average rate $D\avg{\sigma(x)}_+$, where 2D is the dimension of the phase space $\mathcal{C}$ and $\sigma(x)$ is a model-dependent contraction rate per degree of freedom. 

            \medbreak (B) \textit{Reversibility:} There exists an isometry, i.e., a metric preserving map $i: x \rightarrow ix$ in phase space, such that if $t \rightarrow x(t)$ is a solution, then $i(x(-t))$ is also a solution and furthermore $i^2$ is the identity map.

            \medbreak (C) \textit{Chaoticity:} The chaotic hypothesis holds allowing the system ($\mathcal{C}, S$) to be treared as a transitive Anosov system.

            \medbreak\noindent Then, the probability $P_{\tau}(Y)$ that the total entropy production over a time interval $\tau$ takes the value $D\tau\avg{\sigma(x)}_+Y$ satisfies the large-deviation principle:
            \begin{equation}
                \label{eq:fluctuation}
                \ln{\frac{P_{\tau}(Y)}{P_{\tau}(-Y)}}  \;\;\; \myeq \;\;\; D\tau\avg{\sigma(x)}_+Y + \mathcal{O}(1),
            \end{equation}
            where we recall that D is half the dimension of the phase space $\mathcal{C}$.
            
           \section{GCFT and large deviation rescaling} 
                If the GCFT holds, both the symmetry and the derivative of the large deviation function for entropy production are constrained. Specifically, by substituting the form given in Eq.~\ref{eq:LargeDevInf} into Eq.~\ref{eq:fluctuation}, we obtain:
                \begin{equation}
                    \zeta(Y)-\zeta(-Y)=D\avg{\sigma(x)}_+Y.
                    \label{eq:oufzeta}
                 \end{equation}
                Thus, the derivative of the large deviation function at the origin is $D\avg{\sigma(x)}_+/2$, establishing a physical link between the rate of contraction and the large deviations of entropy production.
                                                  
    \subsection{Irreversible or reversible Navier-Stokes equations}
             The classical (irreversible) NS equations, in the incompressible limit by, are given by:
            \begin{eqnarray}
                \partial_t\bm{u} + (\bm{u}\cdot\bm{\nabla})\bm{u}&=& -\bm{\nabla}p + \nu\bm{\Delta}\bm{u} + \bm{f},\label{NS}\\
                \nabla\cdot \bm{u}&=&0,
            \end{eqnarray}
            where $\bm{u}$ is the velocity field, p is the pressure and $\bm{f}$ is a force. For a given forcing, a classical external control parameter of this equation is the Grashoff number:
            \begin{equation}
            Gr=\frac{f_0 L_f^3}{\nu^2},
            \label{NscontrolParameter}
            \end{equation}
            where $L_f$ is a characteristic scale of forcing, defined by $(2\pi/L_f)^2=(1/3)\langle \nabla \bm{f}\cdot \nabla \bm{f}\rangle/\langle \bm{f}\cdot \bm{f}\rangle$ and $f_0=\langle \bm{f}\cdot \bm{f}\rangle^{1/2}$, $\langle .\rangle$ being a spatial average. In the NS equations, the viscosity $\nu$ is constant, and thus even with respect to ${\bm u}$. If one applies the time reversal symmetry $t\to -t;\ \bm{u}\to -\bm{u};\ p\to p$, one observes that the viscous term explicitly breaks the symmetry, resulting in forced irreversibility even in the presence of a time-reversal-symmetric force.

            \medbreak\noindent Time reversal symmetry can actually be restored~\cite{gallavotti1995dynamical} by constructing a time-dependent viscosity  $\nu_r[\bm{u}]$ in such a way as to conserve the total kinetic energy of the system~\cite{shukla2019phase, costa2023reversible, CostaRNS24}. The resulting system of equations is then given by:
            \begin{eqnarray}
                \partial_t\bm{u} + (\bm{u}\cdot\bm{\nabla})\bm{u}&=& -\bm{\nabla}p + \nu_r[\bm{u}]\bm{\Delta}\bm{u} + \bm{f},\label{RNS}\\
                \nabla\cdot \bm{u}&=&0,\\
                \nu_r[\bm{u}] &=& \frac{\int_{\mathcal{D}}\bm{f}\cdot\bm{u} \, \diff\bm{x}}{\int_{\mathcal{D}}\norm{\bm{\nabla}\times\bm{u}}^2\diff\bm{x}}. \label{eq:nu}
            \end{eqnarray}
             For such a system, if the forcing is time-reversal symmetric, the reversible viscosity is odd with respect to ${\bm u}$, ensuring that the viscous term is now time-reversal symmetric, making the RNS equations reversible. In the RNS framework, the viscosity is no longer a constant, so that one needs to build a new external control parameter. Since the total kinetic energy $E=\langle \bm{u}\cdot \bm{u}\rangle/2$ is conserved, the relevant control parameter for a given forcing is the efficiency ${\cal E}$~\cite{CostaRNS24}, a measure of the total kinetic energy stored in the system over the injected energy:
            \begin{equation}
                {\cal E}=\frac{E}{L_f f_0}.
                \label{eq:efficiency}
            \end{equation}
            
            \medbreak\noindent This efficiency controls the dynamics of the RNS system. In particular, it has already been shown using both direct numerical simulations (DNS)~\cite{shukla2019phase} and projections on Log-lattices (LL-RNS)~\cite{costa2023reversible} that the RNS system exhibits a second order phase transition with control parameter $\mathcal{E}^{-1}$ and order parameter being the total enstrophy $\Omega$ (DNS) or its square root $\sqrt{\Omega}$ (LL-RNS). In the latter case, the use of Log-lattices (see Appendix~\ref{Sec:LL}) allowed for an effective probing of the impact of the resolution $N$ on the phase transition, highlighting a gradual shift from an \textit{imperfect} transition (i.e. with exponents different from mean fields predictions) to a \textit{perfect} one as $N$ increases.

        \subsection{Truncated RNS and NS}

            Both the RNS and  the NS are partial differential equations of infinite dimension. 
            To compute the phase space contraction rate, the equations must be projected and truncated onto a finite set of modes, transforming the PDE into a set of ODEs. In the present case, we use Fourier modes on a finite exponentially spaced grid (projection onto log-lattices) to write the equations as:
            \begin{eqnarray}
                &\partial_t{\hat{u}_i} + \bm{i}k_j \hat{u}_j*\hat{u}_i= -\bm{i} k_i\hat{p} - \mu[\bm{u}] k_j k_j \hat{u}_i + \hat{f}_i, \label{eq:Feq} \\
                &\mu[\bm{u}] = \left\{\begin{matrix}
                \begin{array}{ll}
                    \nu, & \text{for NS,} \\
                    \nu_r[\bm{u}], & \text{for RNS,}
                \end{array}
                \end{matrix}\right. \label{eq:Eqnu}
            \end{eqnarray} 
            \noindent where $\vert {\bm k}\vert \le \Lambda$  and ${\bm k}=(\lambda^{n_x},\lambda^{n_y},\lambda^{n_z})$ (LL). In the following, we omit the hat notation for the Fourier transform, as all computations on LL are performed in Fourier space.
              
    \subsection{Phase space contraction rate in NS and RNS}
    
            \begin{figure}[!htb]
                    \begin{center}
                    \subfloat[\label{fig:Sigma_term_comp}]{\includegraphics[width =.45\columnwidth]{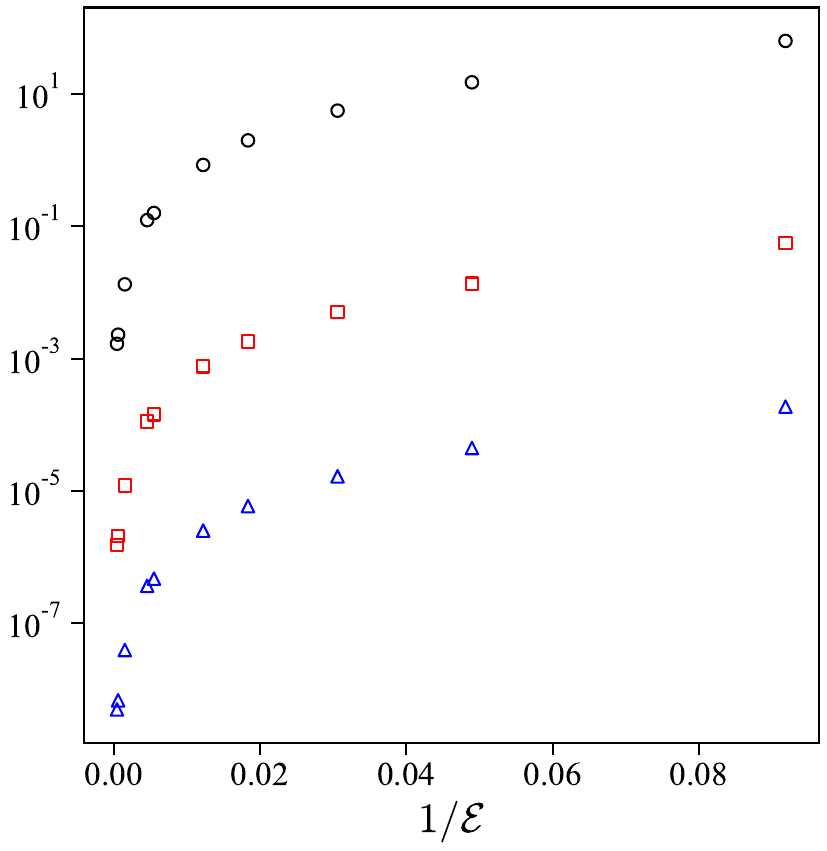}}
                    \subfloat[\label{fig:Mean_sigma}]{\includegraphics[width =.49\columnwidth]{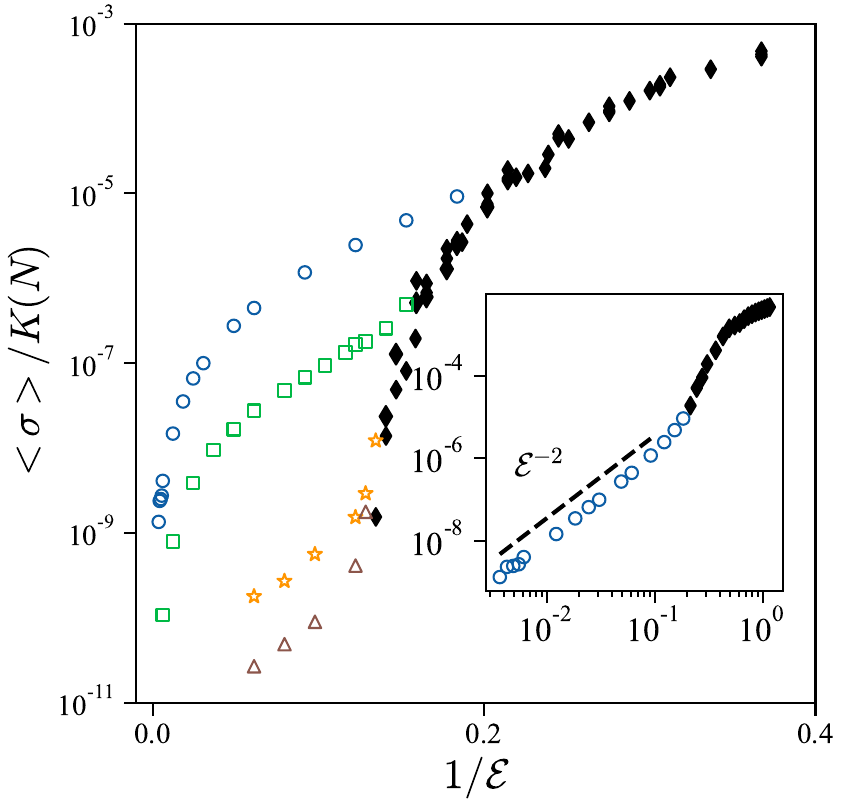}}
                    ~ \caption{(\ref{fig:Sigma_term_comp}) Comparison of time averaged magnitudes of the constitutive terms of $\sigma$ (Eq.~\ref{eq:sigma}) encoded as follow \protect\marksymbol{o}{black}{1.5} $2\avg{\nu_r}K$, \protect\marksymbol{square}{red}{1.5} $2\avg{\nu_r \frac{E_4}{E_2}}$, \protect\marksymbol{triangle}{blue}{1.8} $\avg{\frac{F}{E_2}} $. We observe that $2\avg{\nu_r}K$ is the leading term regardless of the value of $\cal{E}$.
                    (\ref{fig:Mean_sigma}) Rescaled mean phase space contraction rate as a function of the inverse efficiency. Various resolutions are encoded as follow \protect\marksymbol{o}{blueplot}{1.5} $N = 8^3$, \protect\marksymbol{square}{greenplot}{1.5} $N = 12^3$, \resizebox{9pt}{!}{$\color{orangeplot}\smallstar$}\protect  $N = 18^3$, \protect\marksymbol{triangle}{brownplot}{1.8} $N = 20^3$, \protect\diams{black}{1.5} $N = \infty$. For the case \protect\diams{black}{1.5} understand that solutions are always well-resolved regarless of $N$ hence "infinite resolution". The inset emphasizes the power-law behavior - $\avg{\sigma} \propto \mathcal{E}^{-2}$ - in the high efficiency limit.}
                \label{fig:Sigma_def}
                \end{center}
            \end{figure}
            Based on Eqs.~\ref{eq:Feq} \&~\ref{eq:Eqnu}, one can compute the phase space contraction rate $\sigma(\bf{u}) \equiv \sum\limits_{\bm{k}}\frac{\partial{\Dot{u_{\bm{k}}}}}{\partial{u_{\bm{k}}}}$ in both systems, leading to the following expressions:
            \begin{align}
                \sigma_{NS} &= 2\nu K, \label{eq:sigmaNS}\\
                \sigma_{RNS}(\bm u) &= 2\nu_r[\bm u](K - \frac{E_4(\bm u)}{E_2(\bm u)}) + \frac{F(\bm u)}{E_2(\bm u)},
                \label{eq:sigma}
            \end{align}
            where $K = \frac{1}{2}\sum\limits_{\bm{k}} \norm{\bm{k}}^2$, $E_i(\bm u) = \frac{1}{2}\sum\limits_{\bm{k}} \norm{\bm{k}}^{i} \norm{{u_{\bm{k}}}}^2$ and $F(\bm u) = \sum\limits_{\bm{k}} \norm{\bm{k}}^{2} \overline{f_{\bm{k}}}u_{\bm{k}}$. In the following, we omit the subscript for the RNS case when there is no risk of confusion.\
            
             \medbreak\noindent In the NS case, the phase space contraction rate is constant, so that no direct application of the GCFT is possible. While in the RNS case, the phase space contraction rate $\sigma$ is no longer a constant but a fluctuating quantity which in the limit of infnite resolution $N\to\infty$, can be approximated by
            \begin{equation}
                \sigma(\bm u) \approx 2\nu_r[\bm u]K,
                \label{eq:prop}
            \end{equation} 
            as $K$ dominates the other contributions, as highlighted in Figure~\ref{fig:Sigma_term_comp}.
            
            \begin{figure}[!htb]
                \begin{center}
                    \subfloat[\label{fig:Sigma_therm}]{\includegraphics[width =.45\columnwidth]{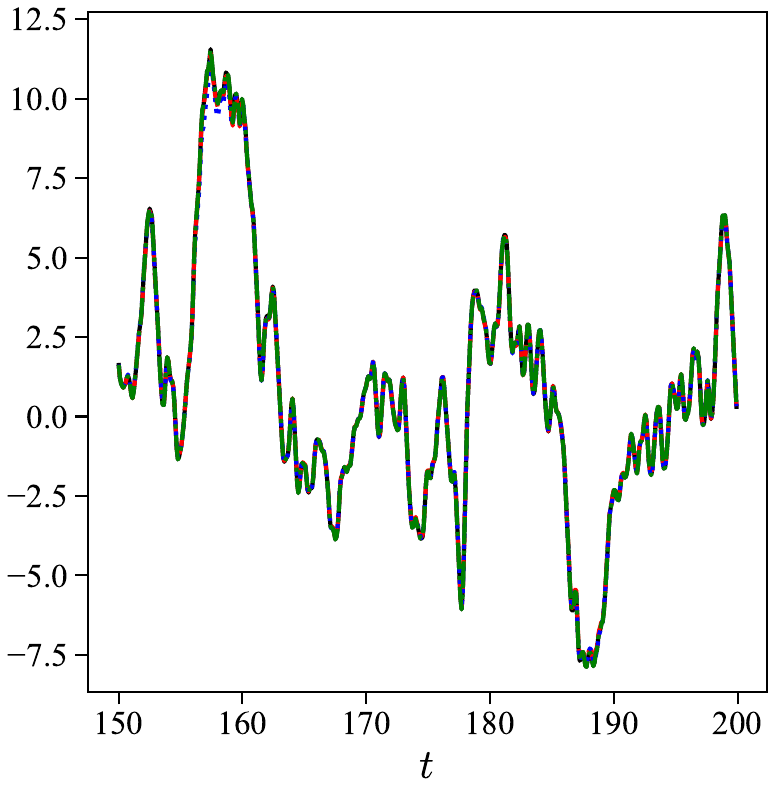}}
                    \subfloat[\label{fig:Sigma_hydro}]{\includegraphics[width =.44\columnwidth]{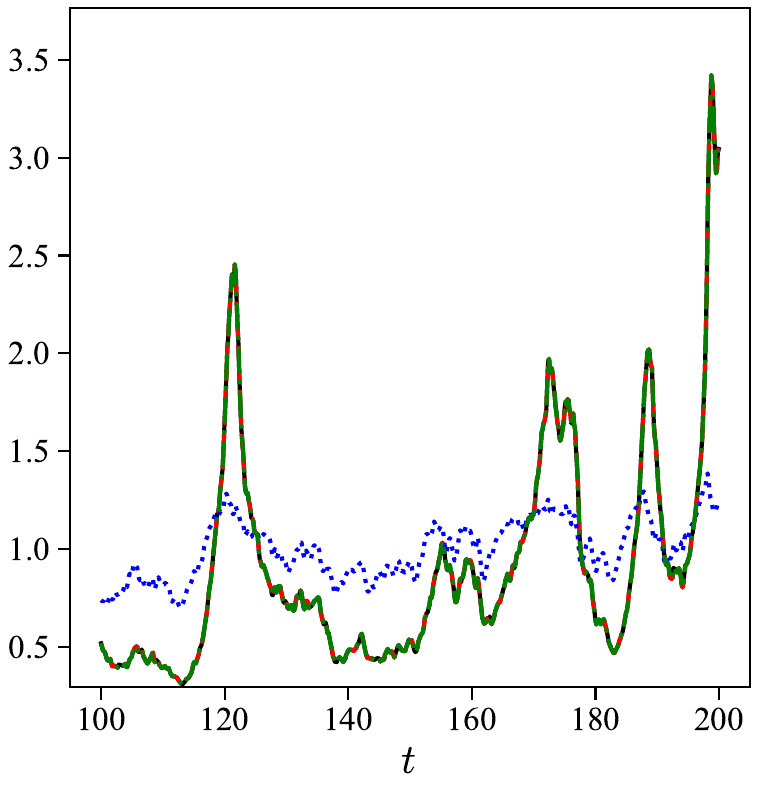}}
                    ~\caption{Relations between different quantities and the rescaled phase space contraction rate. \full{ $\sigma / \avg{\sigma}$}, \textcolor{darkgreen}{\dashdot{ }} $\nu_r / \avg{\nu_r}$, \textcolor{blue}{\dotted{ }} $\epsilon / \avg{\epsilon}$ and \textcolor{red}{\dashed{ }} $X / \avg{X}$, where $X = \epsilon / \Omega$. (\ref{fig:Sigma_therm}) Perfect agreement in the warm phase. (\ref{fig:Sigma_hydro}) The energy injection $\epsilon$ is no longer proportional to the other quantites as $\Omega$ is no longer time-independent for $\cal E \xrightarrow[\cal E > \cal E_*]{} \cal E_*$.}
                    \label{fig:Sigma}
                \end{center}
            \end{figure}
            
            \medbreak\noindent Furthermore, injecting Eq.~\ref{eq:nu} into Eq.~\ref{eq:prop} leads to a relation between the energy injection $\epsilon$, the total enstrophy $\Omega$ and the phase space contraction rate $\sigma$:
            \begin{equation}
                \sigma \propto \epsilon / \Omega.
            \end{equation}
            \noindent This property is illustrated in Figure~\ref{fig:Sigma} through time series of the different quantities related to $\sigma$. In particular, in the high-efficiency regime ${\cal E \gg \cal E_*}$, the enstrophy $\Omega$ is a constant imposed by a thermalized energy spectrum. Consequently, at high efficiency, statistics of $\sigma$ and $\epsilon$ become equivalent in the RNS system (Figure~\ref{fig:Sigma}, full and dotted lines). 
   
    \subsection{Properties of the phase space contraction rate}        
         This  link between $\sigma$, $\Omega$ and $\epsilon$ has two implications:
        \medbreak\noindent (i) In the LL-RNS case, $\sigma$ also undergoes a phase transition (like $\sqrt\Omega$) with control parameter $\mathcal{E}^{-1}$. This is indeed observed in Figure~\ref{fig:Mean_sigma} which presents the averaged and rescaled values of the phase space contraction rate for various resolutions $N$. In particular, one observes that $\avg\sigma / K$ presents two behaviors, a collapse on a master curve for well resolved simulations (Figure~\ref{fig:Mean_sigma}, \protect\diams{black}{1.5}) and resolution-dependent solutions for $\cal{E} > \cal{E}_*$ (where $\mathcal{E}_* \underset{N\rightarrow \infty}{\approx} 7.3$). Note that $\avg{\sigma}$ is always positive which is not guaranteed given that $\nu_r[\bm{u}]$ could be negative (Eq.~\ref{eq:nu}). In addition, the high-efficiency behavior $\avg{\sigma} \propto \mathcal{E}^{-2}$ suggests that $\avg{\sigma} \xrightarrow[\mathcal{E} \rightarrow +\infty]{} 0$ implying perfectly symmetrical PDFs. 
         \medbreak\noindent (ii) In the NS case, we cannot apply GCFT directly to the contraction rate, since it is constant. However, there are other fluctuating quantities such as the energy injection $\epsilon$. According to conjecture by Gallavotti~\cite{gallavotti1996equivalence}, which was numericcally proven on LL in~\cite{costa2023reversible}, the statistical properties of $\epsilon$ in the NS system are equivalent to those of $\epsilon$ in the LL-RNS system. In the LL-RNS case, we found that the statistics of $\epsilon$ are equivalent to those of the entropy production rate in the high-efficiency limit. This provides an incentive to test the GCFT for the energy injection in the case of the NS system at high efficiency.
        \medbreak\noindent In the sequel, we thus first apply the GCFT directly to the phase space contraction $\sigma$ for 3D LL-RNS system before probing its validity in the case of $\epsilon$ for the (irreversible) 3D LL-NS equations.
  
  \section{Application to the Reversible Navier-Stokes equations on log-lattices in 3 dimensions}    
    \subsection{GCFT}
    \subsubsection{Applicability}
       By construction, the \textit{Reversible Navier-Stokes equations} satisfy properties (A)-(C). It is therefore possible to apply the GCFT to the phase space contraction or any related variable.
      
    \subsubsection{Protocol}
      \begin{figure}[!htb]
        \begin{center}
            \subfloat[\label{fig:PDFtau0p0003}]{\includegraphics[width =.49\columnwidth]{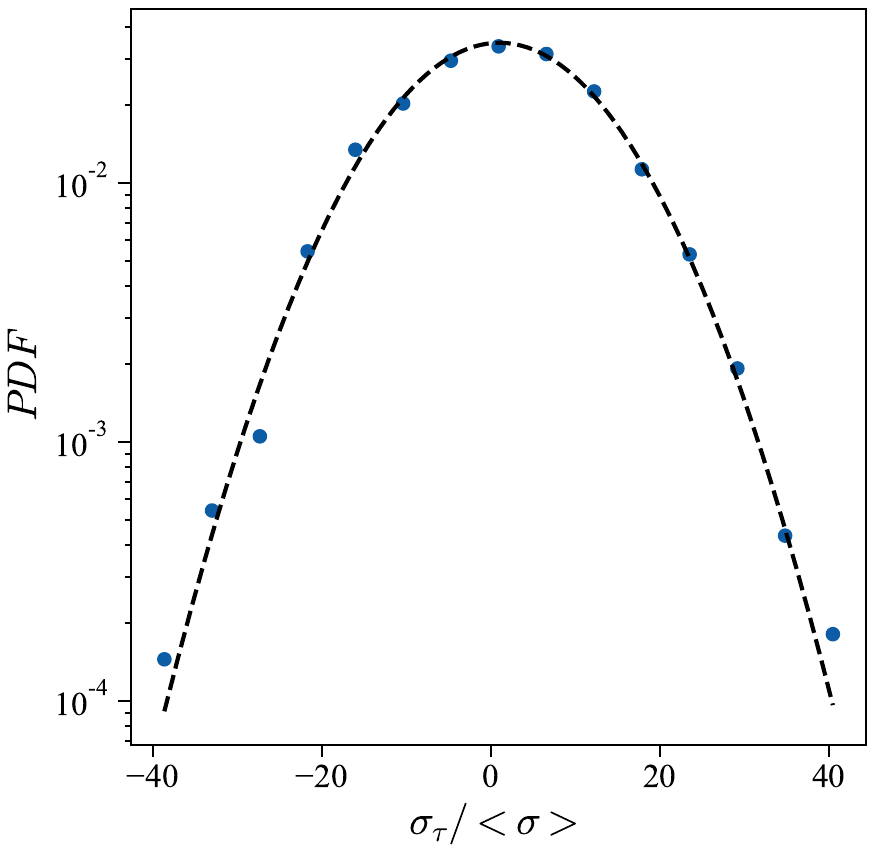}}
            \subfloat[\label{fig:fit0p0003}]{\includegraphics[width =.49\columnwidth]{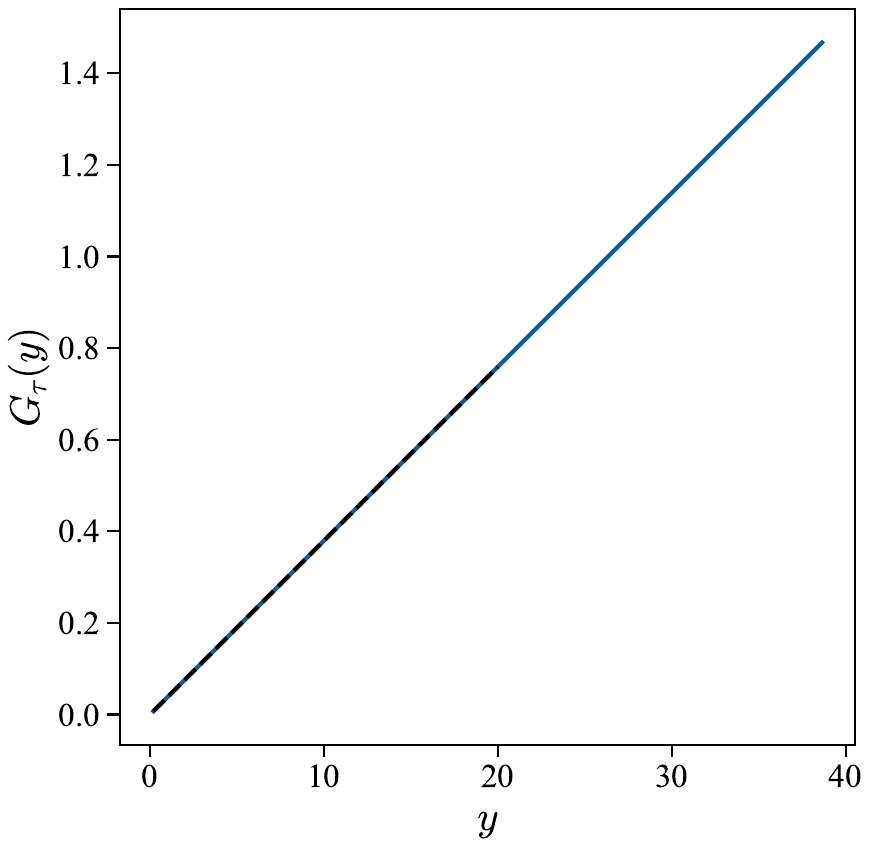}}
            ~\caption{Application of the protocol for $\mathcal{E} = 3266$, $\tau = 4\tau_0$. (\ref{fig:PDFtau0p0003}) Extracted PDF, the dashed line represents the fitted PDF being a skewed-normal distribution. (\ref{fig:fit0p0003}) Plot and fitting of $G_\tau(y)$, $y = \sigma_\tau / \avg{\sigma}$, confirming the linear behavior, on a wide interval around $0$ as predicted by Eq.~\ref{eq:GCFTPi}.}
            \label{fig:PDFfit0p0003}
        \end{center}
    \end{figure}
        In the following we apply the GCFT directly to the phase space contraction rate $\sigma$, given by Eq.~\ref{eq:sigma}. For a given stationary solution of the LL-RNS equations (Eq.~\ref{eq:Feq}), we first compute $\sigma_\tau$ by averaging over a time interval $\tau$:
        \begin{equation}
            \sigma_{\tau}(t) = \frac{1}{\tau}\int_{t}^{t + \tau} \sigma(t')dt'.
        \end{equation}
        Dividing by the mean, we finally obtain the variable $Y = \sigma_\tau / \avg{\sigma}$ and compute its probability distribution function (PDF) $P_\tau(Y)$ by taking long-time statistics over the stationary state. An example is provided in Figure~\ref{fig:PDFfit0p0003}  for $\mathcal{E} \approx 3266$ and $\tau = 4\tau_0$, where $\tau_0$ is the sampling time of the simulation (i.e. the average time between two consecutive saves).
        
       \medbreak\noindent Testing the GCFT  requires the computation the function $ G_\tau(y)$ defined as:
        \begin{equation}
            \label{eq:GCFTPi}
            G_\tau(y) = \frac{1}{\tau}\ln{\frac{P_\tau(Y = y)}{P_\tau(Y = -y)}}.
        \end{equation}
        To avoid statistical limitations, we follow~\cite{ciliberto2004experimental} and proceed in three steps:
        \begin{itemize}
        \item S1: fit the PDF $P_\tau(Y)$ in an interval around $Y=0$~\footnote{The theorem is only applied on a small interval around $0$ where the best fitting is perfectly accurate.} by a suitable distribution ${\tilde P}_\tau(Y)$.
        \item S2: compute the function $G_\tau(y)$ by applying Eq.~\ref{eq:GCFTPi} to ${\tilde P}_\tau(Y)$.
        \item S3: perform a best linear fit of $G_\tau(y)$ following:
        \begin{equation}
        G_\tau(y)=\alpha(\tau)y.
        \label{eq:reffit}
        \end{equation}
        \end{itemize}
      Figure~\ref{fig:PDFfit0p0003} presents the result of the three-step protocol for $\mathcal{E} \approx 3266$ and $\tau = 4\tau_0$. The extracted PDF is fitted using a normal distribution (Figure~\ref{fig:PDFtau0p0003}, dashed lines) which is then used (Figure~\ref{fig:fit0p0003}) to compute $G_\tau(y)$ defined in Eq.~\ref{eq:GCFTPi}. We observe a clear linear behavior on a wide interval, allowing for the extraction of $\alpha(\tau)$ via the slope.
        
       \medbreak To check the validity of the GCFT, we need to vary the value of $\tau$ over a large interval (possibly extending to $\infty$), and check that $\alpha(\tau)$ saturates towards a finite value for large $\tau$, which will then be identified to the mean contraction rate $\avg{\sigma}$. The corresponding computational burden is significant. However, thanks to the projection on LL, we are able to perform extremely long simulations ($t > 5000s$) using small time steps ($dt \approx 5e-3$). This allows us to scan values of $\tau$ ranging from $\tau_0$ to $90\tau_0$.  

      \subsubsection{Fitting functions}
           \begin{figure}[!htb]
            \begin{center}
                \subfloat[\label{fig:PDF0p0003}]{\includegraphics[width =.49\columnwidth]{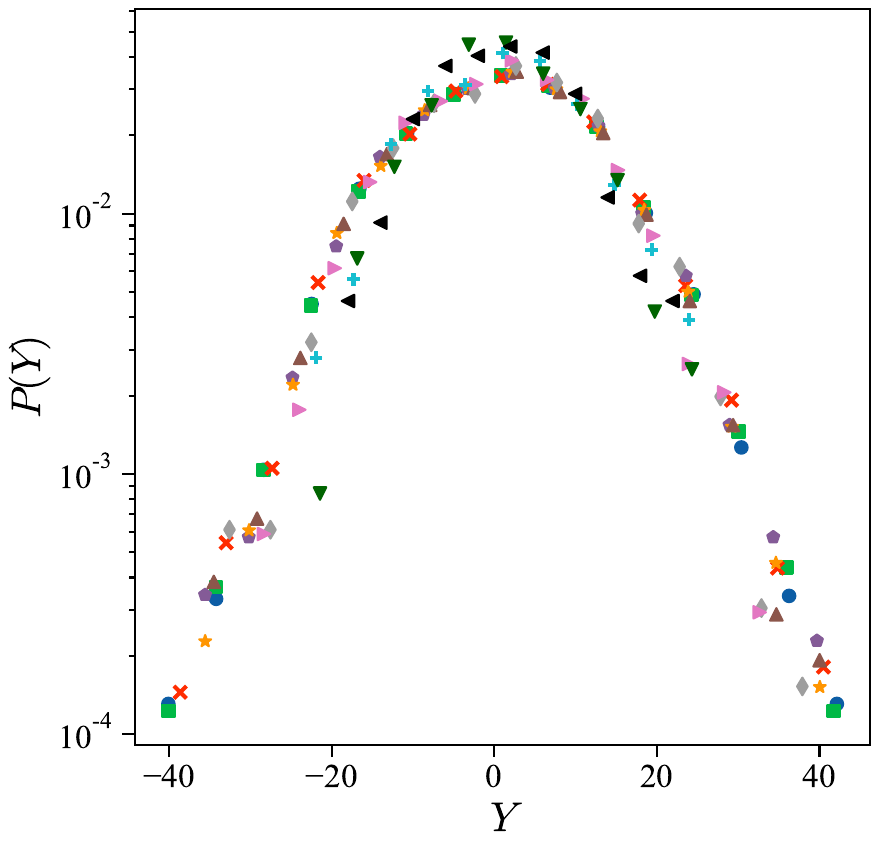}}
                \subfloat[\label{fig:alpha0p0003}]{\includegraphics[width =.49\columnwidth]{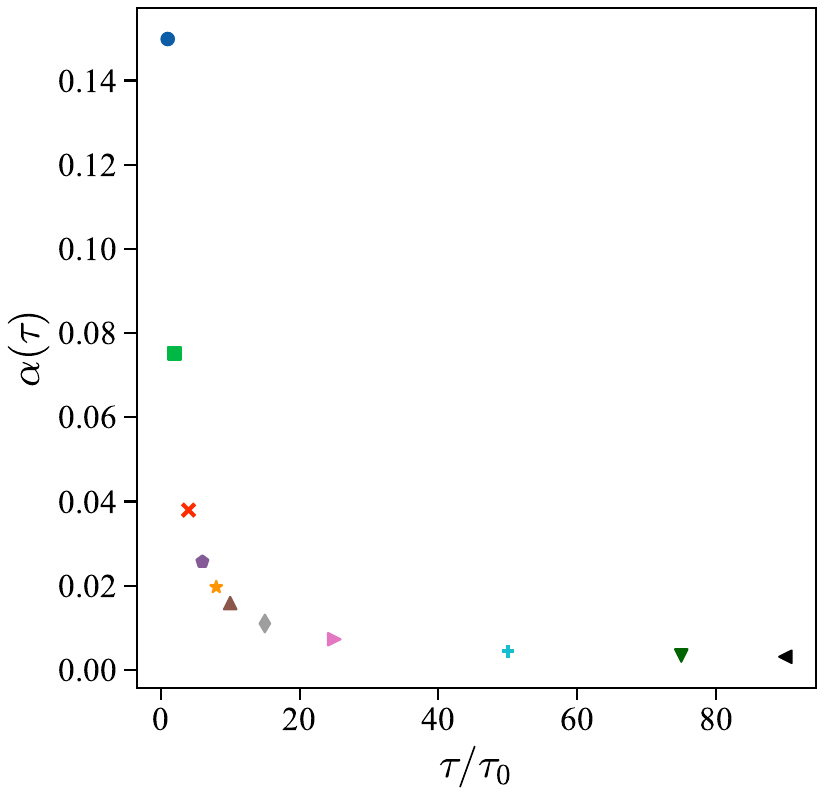}}\\
                \subfloat[\label{fig:PDF0p15}]{\includegraphics[width =.49\columnwidth]{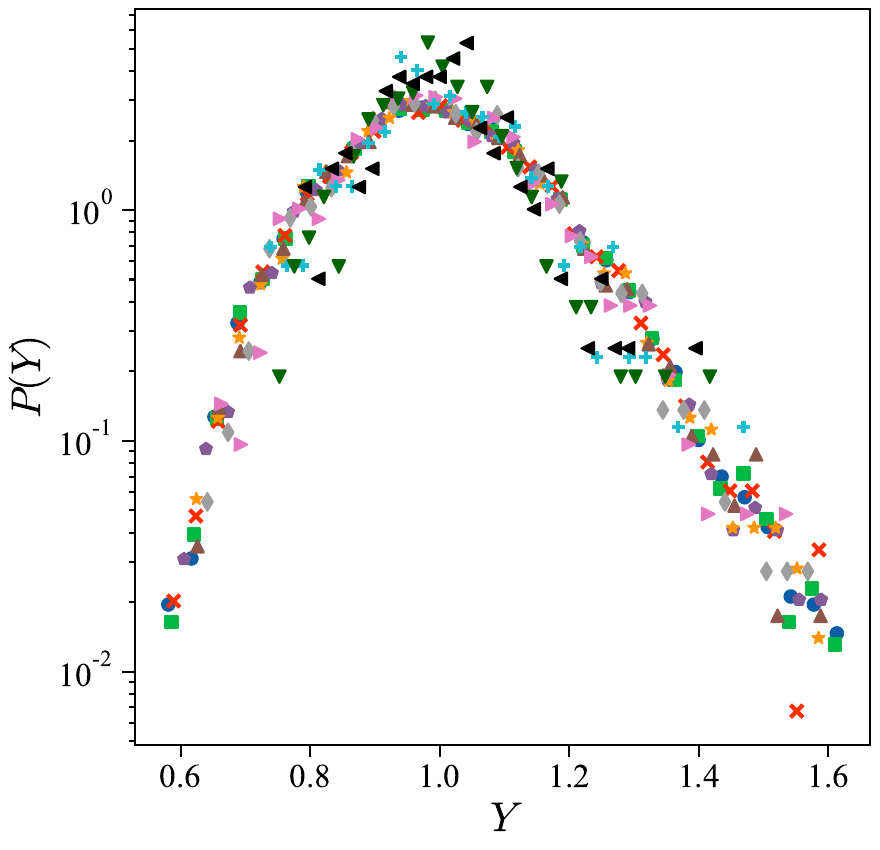}}
                \subfloat[\label{fig:alpha0p15}]{\includegraphics[width =.49\columnwidth]{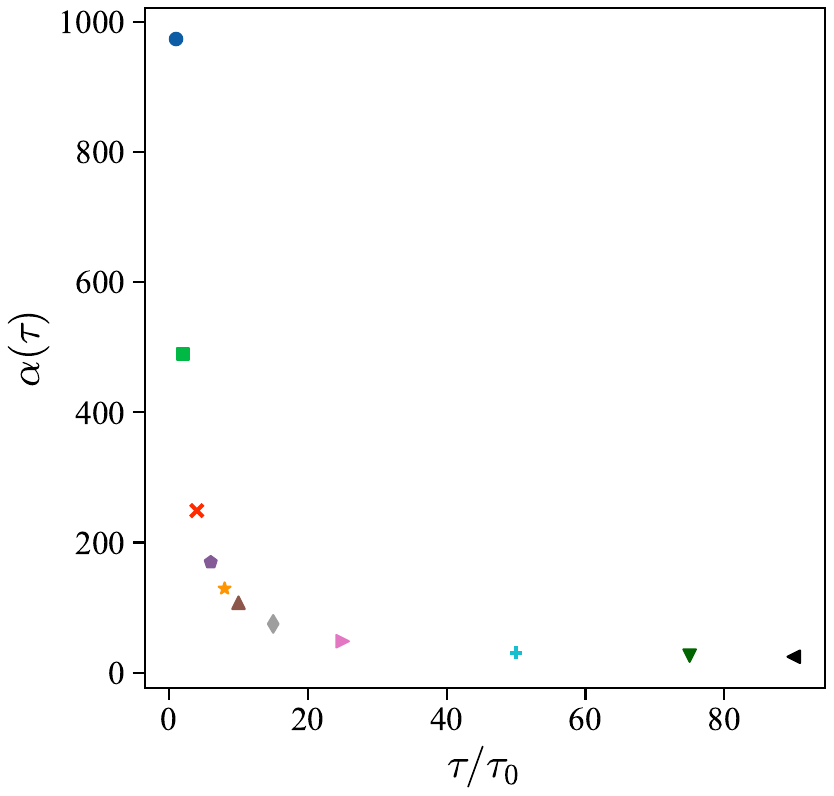}}
                ~\caption{Application of the protocol to several values of $\tau$, results are coded using different symbols and colors. \underline{Top row:} $\mathcal{E} \approx 3266$ (\ref{fig:PDF0p0003}) PDFs of $Y = \sigma_\tau / \avg{\sigma}$.(\ref{fig:alpha0p0003}) extracted values of $\alpha(\tau)$. 
                \underline{Bottom row:} $\mathcal{E} \approx 10.9$ (\ref{fig:PDF0p15}) PDFs of $Y = \sigma_\tau / \avg{\sigma}$. (\ref{fig:alpha0p15}) extracted values of $\alpha(\tau)$.
                In both case $\alpha(\tau)$ saturates at a constant value in the limit of large $\tau$ confirming the prediction of Eq.~\ref{eq:fluctuation}.}
                \label{fig:GCFT0p0003}
            \end{center}
        \end{figure}
            Figures~\ref{fig:PDF0p0003} and~\ref{fig:PDF0p15} show the PDFs of the rescaled phase contraction rate for different values of $\tau$ and two different efficiencies. These figures highlight the dependence of the statistics on these parameters. There are essentially two cases: 

            \begin{enumerate}
                \item For high efficiency, statistics at any $\tau$ can be very well fitted by a normal distribution, with mean $1$ and variance $ s_\tau^2 $: 
                \begin{equation*}
                    \tilde{P}_\tau(y) = \frac{1}{\sqrt{2\pi s_\tau^2}} e^{\frac{-(y - 1)^2}{2s_\tau^2}}.
                \end{equation*}
            
                \item For decreasing efficiency (Figure~\ref{fig:PDF0p15}), the PDFs at any $ \tau $ can be well fitted by skew-normal distributions (positively skewed), defined as:
                \begin{equation*}
                    \tilde{P}_\tau(y) = \frac{1}{\sqrt{2\pi s_\tau^2}} e^{\frac{-(y - \mu)^2}{2s_\tau^2}} \Phi\left(\beta \frac{y - \mu}{s_\tau}\right),
                \end{equation*}
                where $ \Phi(x) = 1 + \text{erf}\left(\frac{x}{\sqrt{2}}\right) $, and $ \text{erf}(x) $ is the error function.
            \end{enumerate}
        
            \noindent Note that the occurrence of negative fluctuations is closely related to the following two factors: 
            \begin{enumerate}
                \item The averaging time $\tau$, as the averaging process tends to decrease the standard deviation of the PDFs. 
                \item The position in the phase space (i.e., the value of $\mathcal{E}$), as the frequency of negative fluctuations decreases in the limit $\mathcal{E} \xrightarrow[\mathcal{E} > \mathcal{E}_*]{} \mathcal{E}_*$ (see Appendix~\ref{Sec:Neg_fluc}).
            \end{enumerate}
            In the absence of negative fluctuations, the latter are recovered through extrapolation of the fitted PDFs.

        \subsubsection{Check of the GCFT symmetry}
\begin{figure}[!htb]
                    \begin{center}
                        \subfloat[\label{fig:Alpha_fit}]{\includegraphics[width =.49\columnwidth]{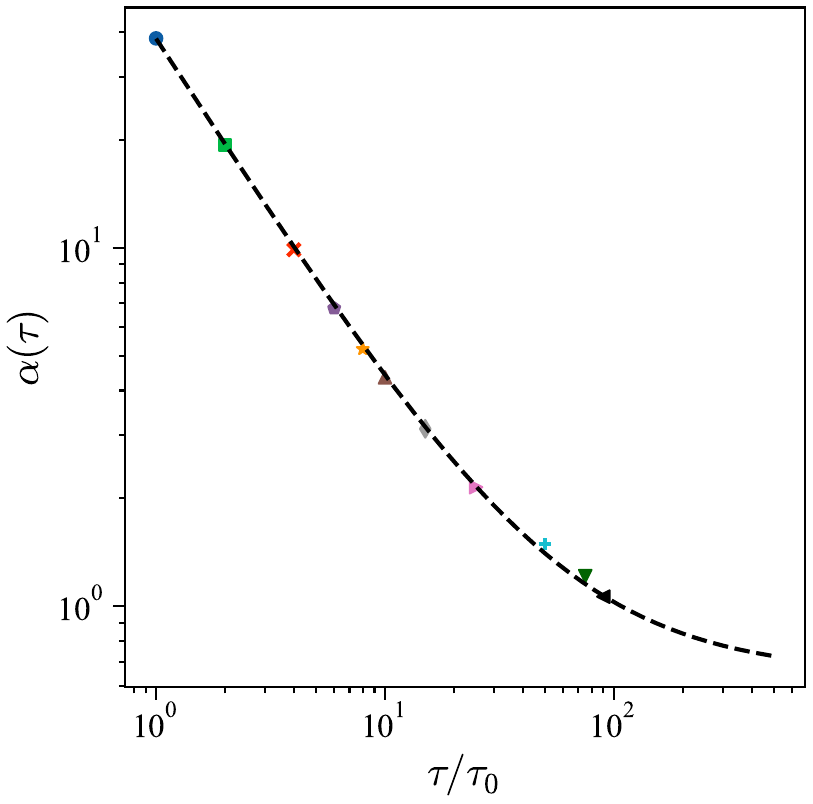}}
                        \subfloat[\label{fig:Alpha_inf}]{\includegraphics[width =.49\columnwidth]{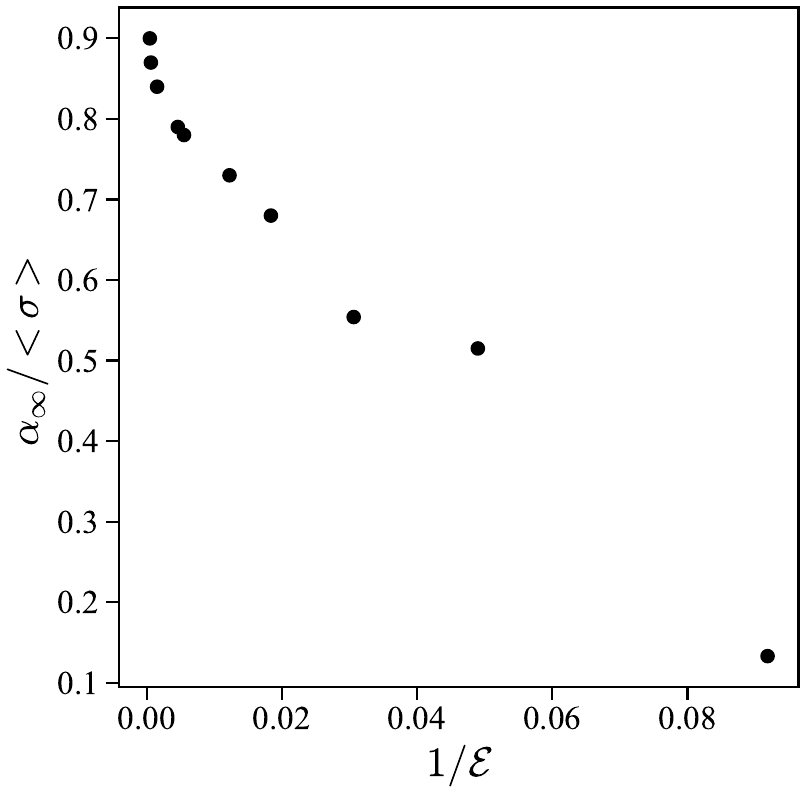}}
                        ~\caption{(\ref{fig:Alpha_fit}) Log-log plot of the extracted slopes $\alpha$ ($\mathcal{E}\approx 81.7$) highlighting a convergence towards a constant value. The dashed line is a fitting $\alpha(\tau) = \alpha_\infty + \,B/\tau$. (\ref{fig:Alpha_inf}) Extracted rate $\alpha_\infty / <\sigma>$ as a function of the control parameter $\mathcal{E}^{-1}$ highlighting decreasing values, from $1$ to $0$, of the rate as the efficiency decreases.}
                        \label{fig:Inf}
                    \end{center}
            \end{figure}

             The fitting procedures (step 1 and 3) allow for an accurate evaluation of $\alpha(\tau)$ for $\tau$ ranging from $\tau_0$ to $90\tau_0$. Corresponding results are presented in Figure~\ref{fig:GCFT0p0003}b and~\ref{fig:GCFT0p0003}d for two different efficiencies.  
             In both cases, $\alpha(\tau)$ decreases as $\tau$ increases before saturating at a constant value $\alpha_\infty(\mathcal{E})$ for large $\tau$, thereby confirming the validity of the GCFT symmetry (Eq.~\ref{eq:fluctuation}). The decrease in both cases can actually be fitted by a simple function, $\alpha(\tau) = \alpha_\infty + \frac{B}{\tau}$. Assuming that the formula holds for any efficiency, we can estimate the value of $\alpha_\infty$ even for cases where saturation occurs at $\tau> 90\tau_0$, which is the maximal time we can afford in our simulations. An example is shown in Figure~\ref{fig:Alpha_fit}, for $\mathcal{E}\approx 81.7$. Applying this technique to all available data, we obtain the behaviour of $\alpha_\infty$ as a function of $\mathcal{E}$. We defer further discussion of this behaviour to the next Section.
         \subsection{Dimension of the attracting set as a function of efficiency}
            \label{Sec:Attractor}
            We have shown in previous sections that for 3D LL-RNS (i) a large deviation function $\zeta(y)$ can be derived for the phase space contraction rate, and (ii) that this large deviation function obeys the GCFT symmetry condition $\zeta(y)-\zeta(-y)=y$. The GFCT theorem further constrains the constant $\alpha_\infty =\avg{\sigma}$.
            \medbreak\noindent The rate $\alpha_\infty =\avg{\sigma}$ as a function of $\mathcal{E}$ is provided in Figure~\ref{fig:Alpha_inf}. At high efficiency, one indeed observes values close to 1 that decreases upon increasing $\mathcal{E}$. To understand such variations, we can apply a remark from~\cite{gallavotti2020ensembles}, regarding the possibility of "partial" reversibility.  
            
            \medbreak\noindent This concept is based on the idea that, although a violation of time reversal symmetry may be observed at the global level when considering the entire phase space, it may still apply to a restricted set of smaller dimension than the full phase space. In this context, the GCFT could still be applied to a restricted space $\mathcal{A'}$ provided that $\mathcal{IA}' = \mathcal{A}'$ (where $\mathcal{I}$) is a metric-preserving map, as discussed in Section~\ref{sec:GCFT_sec} hypothesis (B)).
            
            \medbreak\noindent Applying the GCFT to the restricted space $\mathcal{A'}$ and combining Eqs.~\ref{eq:fluctuation} \&~\ref{eq:GCFTPi}, we obtain the following relation:
            \begin{equation}
                \lim \limits_{\tau \to +\infty}  \alpha(\tau) = \frac{D_\mathcal{A'}}{D_\mathcal{C}}\avg{\sigma}, 
                \label{eq:Alpha_lim}
            \end{equation}
            where where $D_\mathcal{A'}$ (resp. $D_\mathcal{C}$) corresponds to the dimension of the restricted (resp. full) phase space. 
            
            \medbreak\noindent In this context, the decrease of $\alpha_\infty/\avg{\sigma}$ below one as $\cal E \xrightarrow[\cal E > \cal E_*]{} \cal E_*$ could be the signature of the shrinking of the set on which the time reversal symmetry applies. This observation aligns with the fact that the system converges towards an (irreversible) dissipative dynamics in this limit.
   
    \subsection{Universal large deviation rate}
        \begin{figure}[!htb]
        \begin{center}
            \subfloat[\label{fig:Zeta0p0003}]{\includegraphics[width =.49\columnwidth]{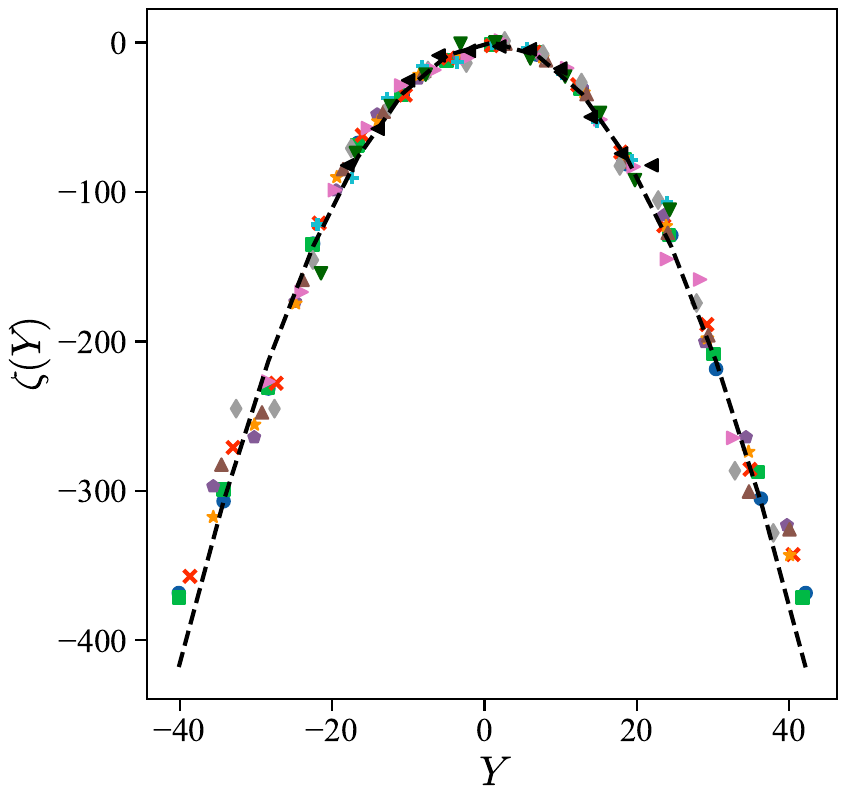}}
            \subfloat[\label{fig:Zeta0p15}]{\includegraphics[width =.49\columnwidth]{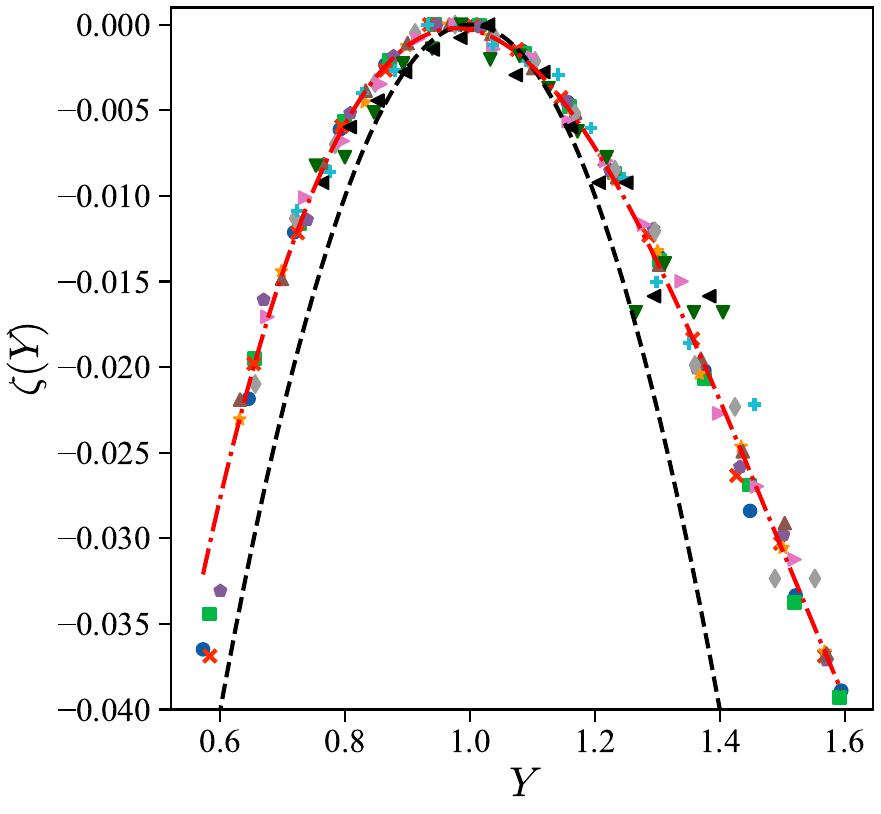}}
            ~\caption{Large deviation rate $\zeta(y)$ introduced in Eq.~\ref{eq:LargeDevGen}, extracted from PDFs presented in Figures~\ref{fig:PDF0p0003} \& \ref{fig:PDF0p15}. (\ref{fig:Zeta0p0003}) High efficiency case associated to normal distributions. (\ref{fig:Zeta0p15}) "Low" efficiency case associated to positively skewed normal distribution. 
            In the low efficiency limit, the rate function closely match the gaussian rate $\zeta_\mathcal{N}$ (Black dashed line). Upon approaching the critical point (i.e. increasing the efficiency) the rate function departs from the gaussian rate and can be estimated by a third order polynomial function $Q$. The colors and symbols encode $\tau/\tau_0$ as in Figure~\ref{fig:alpha0p0003}. $\dashed{}$ $\zeta_\mathcal{N}(y) = -\frac{(y-1)^2}{4}$, $\color{red}\dashdot{}$ $Q \in \mathbb{K}_3[X]$.}
            \label{fig:Zeta}
        \end{center}
    \end{figure}

The shape of the PDF of $Y$ at large $\tau$ is controlled by the large deviation relation (Eq.~\ref{eq:LargeDevInf}), which is valid only at large $\tau$. This can be observed in Figure~\ref{fig:GCFT0p0003}, where a collapse of the PDFs is observed for large $\tau$ but not for small $\tau$.
As remarked by~\cite{ciliberto2004experimental}, there actually exists a universal, finite-time rescaling for $P(Y)$ that generalizes Eq.~\ref{eq:LargeDevInf} and allows for the collapse of all the distributions, even at small values of $\tau$. The rescaling corresponds to:
\begin{equation}
\label{eq:LargeDevGen}
    P_\tau(Y = y) = A_\tau e^{\zeta(y)\tau \alpha(\tau)},
\end{equation}
where $\alpha(\tau)$ is defined in Eq.~\ref{eq:GCFTPi}. This expression can be used to recover the rate function $\zeta$:
\begin{equation}
    \label{eq:zeta}
    \zeta(y) = \frac{\ln P_\tau(Y = y) - \ln A_\tau}{\alpha(\tau) \tau},
\end{equation}
In particular, one has (see Appendix~\ref{Sec:Neg_fluc}) for the normal and skew-normal distributions in the small skewness limit:
\begin{eqnarray}
    A_\tau = \frac{1}{\sqrt{2\pi s_\tau^2}}, \label{eq:Atau} \\
    s_\tau = \sqrt{\frac{2}{\alpha(\tau)\tau}}, \label{eq:stau} \\
    \zeta_{\mathcal{N}}(y) = -\frac{(y-1)^2}{4}, \label{eq:zetagauss}
\end{eqnarray}
where $s_\tau$ is the standard deviation of the distribution.

\medbreak\noindent Figure~\ref{fig:Zeta0p0003} highlights a good match between the extracted rate from Eq.~\ref{eq:zeta} and the prediction from Eq.~\ref{eq:zetagauss} (dashed line) in the high efficiency limit. On the other hand, decreasing the efficiency (i.e., approaching the critical point) is associated with an increase in skewness, causing the extracted rate $\zeta$ to depart (Figure~\ref{fig:Zeta0p15}) from the Gaussian rate function $\zeta_\mathcal{N}$ (black dashed line). In particular, our data are compatible with a polynomial function $Q$ of order 3~\footnote{Note that this expression can only be used in the range where $Q(y) \leq 0$, as the rate function must be negative.} (red dash-dotted line). Quite remarkably, the collapse also holds for the high efficiency case, where negative fluctuations are extrapolated from the fitted PDFs. Note that in both cases, the extracted rate function respects the condition $\zeta'(0) = 1/2$ (see Eq.~\ref{eq:oufzeta}).
\begin{figure}[!htb]
        \begin{center}
            \subfloat[\label{fig:PDF1e-11}]{\includegraphics[width =.48\columnwidth]{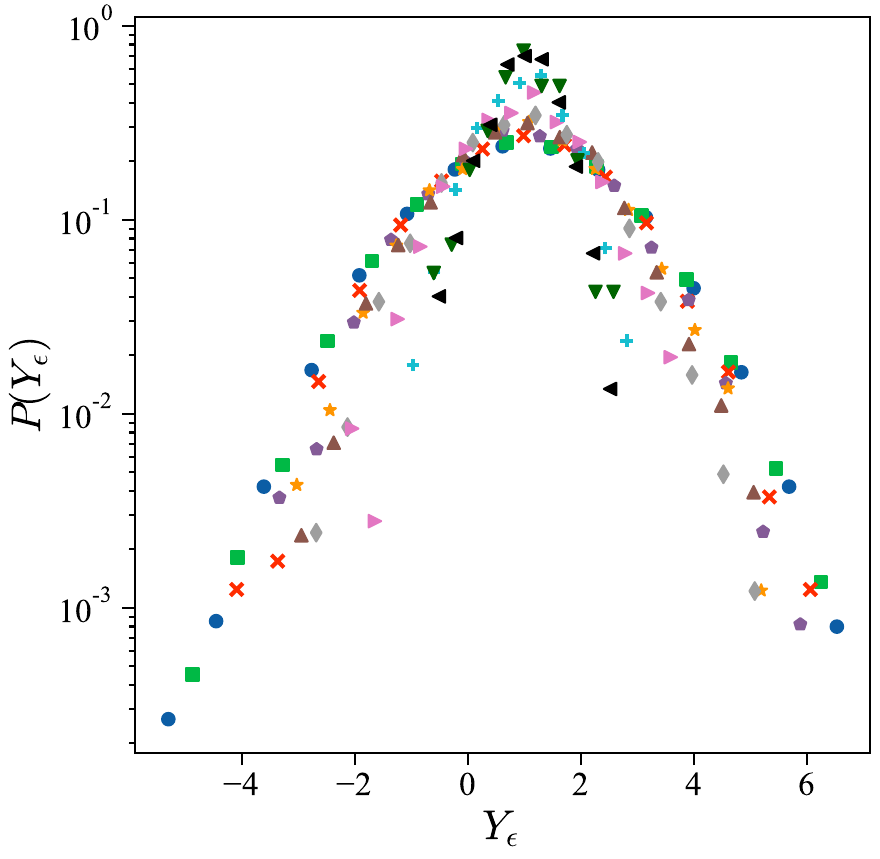}}
            \subfloat[\label{fig:PDF1e-07}]{\includegraphics[width =.48\columnwidth]{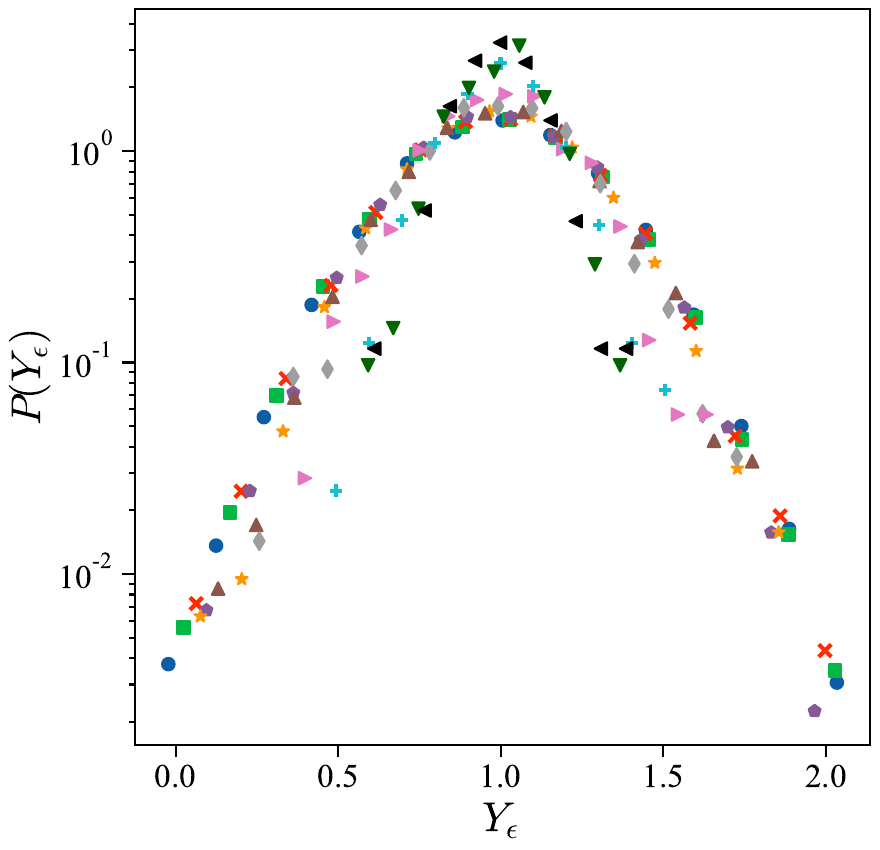}}\\
            \subfloat[\label{fig:Sinai1e-11}]{\includegraphics[width =.48\columnwidth]{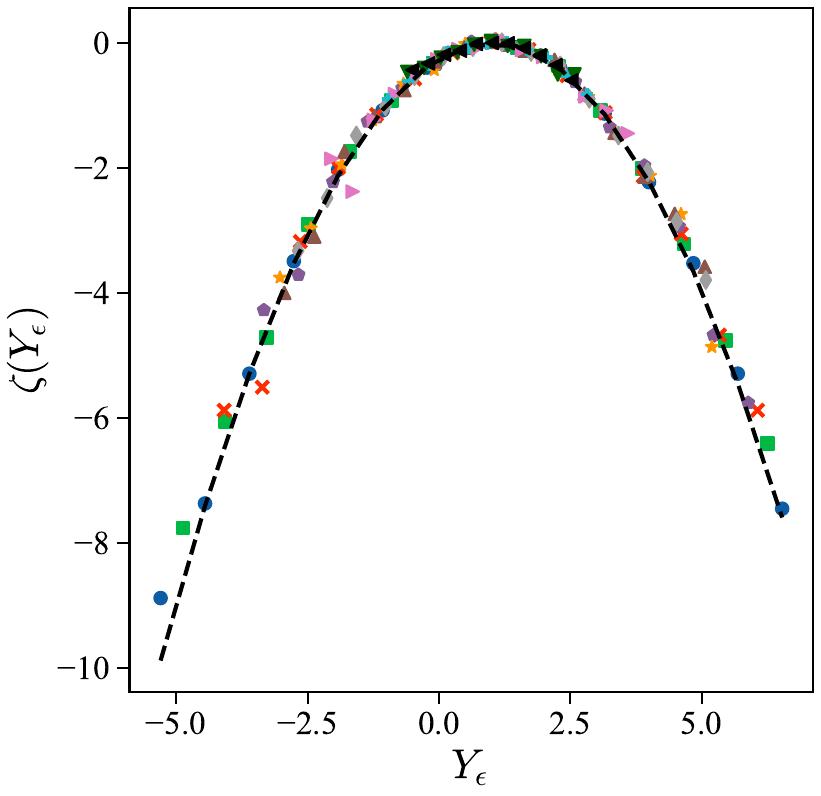}}
            \subfloat[\label{fig:Sinai1e-07}]{\includegraphics[width =.492\columnwidth]{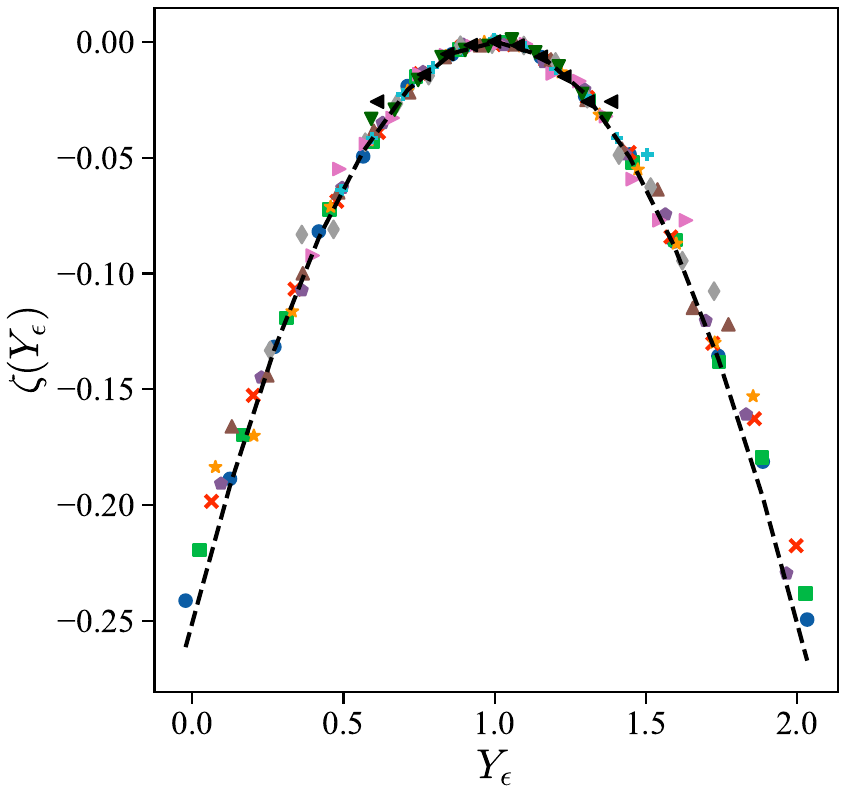}}
            ~\caption{Application of the GCFT to the LL-NS equations in term of injection i.e $Y_\epsilon = \epsilon_\tau / \avg{\epsilon}$. Color and symbols encode $\tau / \tau_0$ as in Figure~\ref{fig:GCFT0p0003}. Two different viscosities are used (\ref{fig:PDF1e-11}) \& (\ref{fig:Sinai1e-11}) $\nu = 1e-11$, $\mathcal{E} \approx 237$. (\ref{fig:PDF1e-07}) \& (\ref{fig:Sinai1e-07}) $\nu = 1e-07$, $\mathcal{E} \approx 40$.
            \underline{Top row:} Extracted PDFs of $Y_\epsilon$. \underline{Bottom row:} Large deviation rates $\zeta(y)$ introduced in Eq.~\ref{eq:LargeDevGen}, extracted from the above PDFs.}
            \label{fig:NSGCFT}
        \end{center}
    \end{figure}
\section{Extending the results to Navier-Stokes}
        We now investigate the validity of the GCFT for the standard 3D LL-NS equations using the energy injection $\epsilon = \bm{f} \cdot \bm{u}$, which might exhibit negative fluctuations. As previously mentioned, this choice is motivated by the observation that in the high efficiency limit, $\sigma / \avg{\sigma} = \epsilon / \avg{\epsilon}$ for LL-RNS, and by the Gallavotti equivalence conjecture~\cite{gallavotti1995dynamical, costa2023reversible} between NS and RNS systems.

        \medbreak\noindent For extremely small viscosities, the system is affected by strong Galerkin truncation effects, making it effectively reversible. It is then no surprise to observe behaviors similar to the LL-RNS case. Indeed, the energy injection presents negative fluctuations, the frequency of which decreases for increasing $\tau$ (Figure~\ref{fig:PDF1e-11}; $\nu = 1e-11$, $\avg{\mathcal{E}} \approx 237$) until their complete disappearance. By increasing the viscosity, one reaches statistically steady states of decreasing total kinetic energy $E$, associated with decreasing - averaged - efficiency $\avg{\mathcal{E}}$. Hence, we expect - from the LL-RNS results - to observe fewer and fewer negative fluctuations of $\epsilon$ in the limit$ \avg{\mathcal{E}} \xrightarrow[\avg{\mathcal{E}} > \mathcal{E}_*]{} \mathcal{E}_*$. Such findings would then be in agreement with the observed shrinking of the attracting set discussed in Section~\ref{Sec:Attractor}.

\section{Conclusion}
In this work, we have investigated the Gallavotti-Cohen fluctuation theorem (GCFT) using the projection of fluid equations onto Log-lattices~\cite{campolina2021fluid}. In particular, it has been shown that for the Reversible Navier-Stokes (RNS) system, the phase space contraction rate $\sigma$ is proportional to the reversible viscosity $\nu_r$, which undergoes a phase transition~\cite{costa2023reversible}. Therefore, the phase space contraction rate undergoes the same phase transition, separating two phases: 
\begin{enumerate}
    \item The warm phase (high efficiency limits), where there exists an attracting set $ \mathcal{A'} \subset \mathcal{C}$, eventually smaller than the phase space $\mathcal{C}$, on which time reversal is indeed a symmetry.
    \item The hydrodynamic phase ($\mathcal{E} \ll \mathcal{E}_*$) associated with a rapid increase of the averaged viscosity $\avg{\nu_r}$, implying a decrease in the \textit{viscous time scale} $T_\nu \equiv L^2 / \avg{\nu_r}$. This impedes the study of fluctuations, as only fluctuations over time intervals $\tau \ll T_\nu$ can be effectively studied through large deviations and the fluctuation theorem~\cite{gallavotti2014nonequilibrium}.
\end{enumerate}

\medbreak\noindent Nonetheless, we have shown that the GCFT can effectively be applied to a subset $\mathcal{A'}$ of the phase space $\mathcal{C}$ and that the fluctuations indeed follow a large deviation theory with rate function $\zeta$, which can be estimated from the form of the PDF. Quite surprisingly, the GCFT has been successfully applied to the LL-NS equations, which exhibits irreversible dynamics. This success is most likely due to the fact that dissipation does not occur in the bulk but rather as a boundary effect at large wave numbers. Hence, the GCFT could be applied to a wider range of systems, often more realistic, than the class of reversible systems.

\section{Acknowledgment}
This work received funding through the PhD fellowship programs of the Ecole Normale Sup\'erieure Paris-Saclay, and through the ANR, via the grants ANR TILT grant agreement no.\ ANR-20-CE30-0035, ANR BANG grant agreement no.\ ANR-22-CE30-0025. This work was supported by a French government grant managed by the Agence Nationale de la Recherche France 2030 under the reference ANR-24-RRII-0001.

\section*{Appendices}

    \subsection{Numerical framework - Log-lattices}
        \label{Sec:LL}
         Performing turbulent simulations (i.e., high Reynolds numbers $Re$) comes with significant numerical challenges, as Direct Numerical Simulations (DNS) of the Navier-Stokes Equations (NSE) have a computational complexity of the order $\mathcal{O}(Re^3)$. Introducing time-dependent viscosity further increases the computational cost, requiring an additional step in the numerical resolution of the Reversible Navier-Stokes (RNS) equations, thus exacerbating the numerical burden. Due to these limitations, DNS of both the NS and RNS systems are typically restricted to small resolutions, which hampers the ability to probe many important properties of turbulent flows, particularly at high Reynolds numbers.

        \medbreak In this paper, we propose the use of a recent framework known as Log-Lattices (LL), developed by Campolina and Mailybaev~\cite{campolina2018chaotic, campolina2021fluid}, in which the equations are projected onto a logarithmic grid in Fourier space. In this approach, the modes follow a geometric progression $k_n = k_0 \lambda^n$, where $\lambda$ is the spacing parameter of the lattice. By utilizing this framework, it becomes possible to observe the long-time behavior of flows at high Reynolds numbers. This is especially valuable in the study of Gallavotti-Cohen Fluctuation Theorem (GCFT), as the asymptotic behavior $\tau\to\infty$ is often not attainable in DNS simulations due to the inherent computational costs.

        \subsection{Anosov systems and Fluctuation Theorem}
            \label{Sec:Sinai}
                A differential dynamical system is defined by the existence of a bijection $S_t : \mathcal{C} \rightarrow \mathcal{C}$, $\mathcal{C}$ being the phase space, such that for any initial condition $x_0$, there exists a unique futur at time t:
                \begin{equation}
                    x(t) = S_t(x_0).
                \end{equation}
                Varying t, one obtains a flow in the phase space $\mathcal{C}$ such that~\footnote{A simple example is the Hamiltonian flow of classical mechanic.}:
                \begin{eqnarray}
                    & S_0 = \text{Id}, \\
                    & \forall (t, s) \in \mathbb{R}^2, \text{ } S_t \circ S_s = S_{t+s}.
                \end{eqnarray}

                \medbreak\noindent An \textit{Anosov system}, is a \textit{smooth} dynamical system ($\mathcal{C}, S$) such that the phase space $\mathcal{C}$ can be decomposed as $\mathcal{C} = E_0 \bigoplus E_S\bigoplus E_I$, where $E_0$ is a 1D manifold in the flow direction while $E_S$ (resp. $E_I$) is a stable (resp. unstable) manifold that exponentially contracts (resp. expands) corresponding to negative (resp. positive) Lyapunov exponents.
                
                \noindent In addition, an \textit{Anosov system} is said to be \textit{transitive} if the stable and unstable manifolds are dense in $\mathcal{C}$~\cite{gallavotti1995dynamical}.

                \medbreak\noindent For such systems, the following theorem holds:
                \medbreak\textit{Theorem~\cite{sinai1977lectures}: if a system is a transitive Anosov system, then it admits a Sinai-Ruelle-Bowen (SRB) distribution.}

                 \noindent We recall that a SRB distribution $\mu$ is defined such that~\cite{gallavotti2005srb}:
                 \begin{equation}
                 \label{eq:SRB}
                     \lim_{T\rightarrow\infty} \frac{1}{T}\sum_{i=0}^{T-1}f(S^i(x)) = \int_{\mathcal{C}} \mu(dy)f(y),
                 \end{equation}
                for all smooth function $f$ and all $x \in \mathcal{C}$. In particular, Eq.~\ref{eq:SRB} assesses the existence of the time-average of any macroscopic observable, defined at a point $x$ of the phase space $\mathcal{C}$.

                \medbreak\noindent For a time-reversible, dissipative (i.e $\avg\sigma > 0$, $\sigma$ being the phase space contraction rate), transitive Anosov system, the probability distribution $P_\tau(p)$ of the variable
                \begin{equation}
                    p = \frac{1}{\tau}\int_0^\tau \frac{\sigma(S^tx)}{\avg\sigma}dt,
                \end{equation}
                verifies the large deviation relation $P_\tau(p) = e^{\zeta(p)\tau + \mathcal{O}(1)}$~\cite{sinai1977lectures}, with the symmetry property $\zeta(p) - \zeta(-p) = p\avg{\sigma}$ for all $p$ within the definition domain of $\zeta$.

                \medbreak\noindent This fluctuation theorem although interesting has only been rigorously proven for idealized systems such as Anosov systems. The extension to real systems, made of many particles is therefore non-trivial. In the following, we tackle the subject of the \textit{Gallavotti-Cohen Fluctuation Theorem}, obtained through less restrictive hypothesis.

    \subsection{On the existence of negative fluctuations of the phase rate contraction rate}
    \label{Sec:Neg_fluc}
    \noindent Suppose that the sampled viscosity $\avg{\nu_r}_\tau$ ($\propto \sigma_\tau$) follows a skew-normal distribution $(\mu, s_\tau, \beta)$, then one obtains that:
    \begin{eqnarray}
        &\delta &= \frac{\beta}{\sqrt{1 + \beta^2}}, \label{eq:delta}\\
        &1 &= \mu + s_\tau\delta\sqrt{\frac{2}{\pi}}, \label{eq:Mean} \\
        &\alpha(\tau)\tau &= \frac{2\mu}{s_\tau^2}(1+\beta^2), \label{eq:alphamu}\\
        &Var(\frac{\nu_r}{\avg{\nu_r}})f(\tau) &= s_\tau^2(1 - \frac{2\delta^2}{\pi}) \label{eq:Var}.
    \end{eqnarray}
    In particular, the expression~\ref{eq:alphamu} has been obtained using an approximation~\cite{abramowitz1968handbook} of the error function around $0$. Table.~\ref{tab:Erroralpha} presents the percentage of error between the extracted slopes and the approximated expression highlighting a good match. 

\begin{table}[h]
    \centering
    \resizebox{1.\columnwidth}{!}{
    \begin{tabular}{|c|c||c|c|c|c|c|c|c|c|c|c|c|} \hline
        \multicolumn{2}{|c|}{\multirow{2}{*}{}} &                           
        \multicolumn{11}{|c|}{$\tau/\tau_0$ }  \\ \cline{3-13} 
        \multicolumn{2}{|c|}{}   & 1 & 2 & 4 & 6 & 8 & 10 & 15 & 25 & 50 & 75 & 90\\ \cline{1-13}
        \hline \hline
        \multirow{3}{*}{$\mathcal{E}$} & 2177.7  & 3e-8 & 1.9e-8 & 4.9e-13 & 2e-10 & 2.3e-9 & 4e-9 & 6.1e-10 & 3.7e-10 & 6.7e-14 & 2.5e-14 & 8.2e-10 \\
        \cline{2-2}
        & 181.5 & 6.4e-3 & 1.2e-3 & 1.9e-3 & 1.6e-2 & 2.6e-2 & 4.5e-2 & 4.4e-2 & 2.3e-2 & 1.7e-1 & 3.7e-1 & 3.6e-1 \\
         \cline{2-2}
         & 10.9 & 1.92 & 1.92 & 1.89 & 1.87 & 1.87 & 1.82 & 1.77 & 1.64 & 1.39 & 1.13 & 1.12  \\
        \hline
    \end{tabular}}
    \caption{Percentage of error $\frac{|\alpha(\tau)\tau - \alpha_{\mathcal{SN}}|}{\alpha(\tau)\tau}$ highlighting a good match between the extracted slopes $\alpha(\tau)\tau$ and the approximated (from skew-normal distribution) expression $\alpha_{\mathcal{SN}}(\beta, \mu,s_\tau)\tau = \frac{2\mu}{s_\tau^2}(1+\beta^2)$. We recall that $\tau$ corresponds to the averaging time while $\tau_0$ is sampling time of the simulations.}
    \label{tab:Erroralpha}
\end{table}
\noindent Negative values of the viscosity can be observed when $s_\tau \geq \mu$, i.e., when (Eq.~\ref{eq:Mean}) $s_\tau \geq \left(1 + \delta \sqrt{\frac{2}{\pi}}\right)^{-1}$. It has already been shown~\cite{costa2023reversible} that the RNS system exhibits a phase transition with the control parameter $\Rr$. Combining these results with Eq.~\ref{eq:Var} implies that $s_\tau^2 \sim C \left(1 - \frac{2\delta^2}{\pi}\right)^{-1} (\Rr - \Rrstar)^{\gamma}$. Hence, one obtains the following equation:
    \begin{equation}
        \left(1 + \delta \sqrt{\frac{2}{\pi}}\right)^{-1} = \sqrt{C_\tau \left(1 - \frac{2\delta^2}{\pi}\right)^{-1}} (\Rrstar - \Rr)^{\gamma/2}, \quad \forall \mathcal{E} > \mathcal{E}_*.
    \end{equation}
    \noindent In particular, near the critical point, one observes a maximum skewness, i.e., $\beta \gg 1$, leading to $\delta \approx 1$, so that the above equation now reads:
    \begin{equation}
        \left(1 + \sqrt{\frac{2}{\pi}}\right)^{-1} = \sqrt{C_\tau \left(1 - \frac{2}{\pi}\right)^{-1}} (\Rrstar - \Rr)^{\gamma/2}.
        \label{eq:Bounds}
    \end{equation}
    \noindent From Eq.~\ref{eq:Bounds}, one recovers a second critical value $\Rr^-$ of the control parameter:
    \begin{equation}
    \begin{split}
        \Rr^- &= \Rrstar - \,\Delta_\tau,\\
        \Delta_\tau  &= \left[\sqrt{\frac{C_\tau}{1 + \frac{2}{\pi}}} \left(1 + \sqrt{\frac{2}{\pi}}\right)\right]^{2/\gamma} > 0.
    \end{split}
    \end{equation}
    \noindent For $\forall \mathcal{E} \geq \mathcal{E}^-$ one observes negative values of the viscosity, whose fluctuations can be obtained through the GCFT.

    \medbreak\noindent In addition, the asymptotic behavior for $\mathcal{E} \gg \mathcal{E}_*$ can be obtained for the set of equations~\ref{eq:delta}-\ref{eq:Var} by taking the limit $\beta \to 0$ (i.e., $\delta \to 0$), recovering a normal distribution for the viscosity (and so for the phase space contraction rate), which indeed leads to Eqs.~\ref{eq:Atau} \&~\ref{eq:stau}.

\providecommand{\noopsort}[1]{}\providecommand{\singleletter}[1]{#1}%


\begin{thebibliography}{10}

\bibitem{sinai1977lectures}
Ya~G Sinai.
\newblock Lectures in ergodic theory.
\newblock {\em Lecture notes in Mathematics. Princeton University Press, Princeton}, 4:1121--1131, 1977.

\bibitem{evans1993probability}
Denis~J Evans, Ezechiel Godert~David Cohen, and Gary~P Morriss.
\newblock Probability of second law violations in shearing steady states.
\newblock {\em Physical review letters}, 71(15):2401, 1993.

\bibitem{gallavotti1995dynamical}
Giovanni Gallavotti and Ezechiel Godert~David Cohen.
\newblock Dynamical ensembles in stationary states.
\newblock {\em Journal of Statistical Physics}, 80:931--970, 1995.

\bibitem{ruelle1995measures}
David Ruelle.
\newblock Measures describing a turbulent flow.
\newblock {\em Turbulence, Strange Attractors, and Chaos}, 16:223, 1995.

\bibitem{ciliberto1998experimental}
S~Ciliberto and C~Laroche.
\newblock An experimental test of the gallavotti-cohen fluctuation theorem.
\newblock {\em Le Journal de Physique IV}, 8(PR6):Pr6--215, 1998.

\bibitem{ciliberto2004experimental}
Sergio Ciliberto, Nicolas Garnier, S~Hernandez, C{\'e}drick Lacpatia, J-F Pinton, and G~Ruiz Chavarria.
\newblock Experimental test of the gallavotti--cohen fluctuation theorem in turbulent flows.
\newblock {\em Physica A: Statistical Mechanics and its Applications}, 340(1-3):240--250, 2004.

\bibitem{gallavotti1996equivalence}
Giovanni Gallavotti.
\newblock Equivalence of dynamical ensembles and navier-stokes equations.
\newblock {\em Physics Letters A}, 223(1-2):91--95, 1996.

\bibitem{margazoglou2022nonequilibrium}
Georgios Margazoglou, Luca Biferale, Massimo Cencini, Giovanni Gallavotti, and Valerio Lucarini.
\newblock Nonequilibrium ensembles for the three-dimensional navier-stokes equations.
\newblock {\em Physical Review E}, 105(6):065110, 2022.

\bibitem{shukla2019phase}
Vishwanath Shukla, B{\'e}reng{\`e}re Dubrulle, Sergey Nazarenko, Giorgio Krstulovic, and Simon Thalabard.
\newblock Phase transition in time-reversible navier-stokes equations.
\newblock {\em Physical Review E}, 100(4):043104, 2019.

\bibitem{costa2023reversible}
Guillaume Costa, Amaury Barral, and B{\'e}reng{\`e}re Dubrulle.
\newblock Reversible navier-stokes equation on logarithmic lattices.
\newblock {\em Physical Review E}, 107(6):065106, 2023.

\bibitem{CostaRNS24}
Guillaume Costa, Amaury Barral, Quentin Pikeroen, and Bérengère Dubrulle.
\newblock Behind the mirror: the hidden dissipative singular solutions of ideal reversible fluids on log-lattice.
\newblock {\em arXiv preprint arXiv:2508.10659}, 2025.

\bibitem{campolina2021fluid}
Ciro~S Campolina and Alexei~A Mailybaev.
\newblock Fluid dynamics on logarithmic lattices.
\newblock {\em Nonlinearity}, 34(7):4684, 2021.

\bibitem{gallavotti2020ensembles}
Giovanni Gallavotti.
\newblock Ensembles, turbulence and fluctuation theorem.
\newblock {\em The European Physical Journal E}, 43:1--10, 2020.

\bibitem{gallavotti2014nonequilibrium}
Giovanni Gallavotti.
\newblock {\em Nonequilibrium and irreversibility}.
\newblock Springer, 2014.

\bibitem{campolina2018chaotic}
Ciro~S Campolina and Alexei~A Mailybaev.
\newblock Chaotic blowup in the 3d incompressible euler equations on a logarithmic lattice.
\newblock {\em Physical review letters}, 121(6):064501, 2018.

\bibitem{gallavotti2005srb}
Giovanni Gallavotti.
\newblock Srb distribution for anosov maps.
\newblock In {\em Chaotic Dynamics and Transport in Classical and Quantum Systems: Proceedings of the NATO Advanced Study Institute on International Summer School on Chaotic Dynamics and Transport in Classical and Quantum Systems Carg{\`e}se, Corsica 18--30 August 2003}, pages 87--105. Springer, 2005.

\bibitem{abramowitz1968handbook}
Milton Abramowitz and Irene~A Stegun.
\newblock {\em Handbook of mathematical functions with formulas, graphs, and mathematical tables}, volume~55.
\newblock US Government printing office, 1968.

\end{thebibliography}
\end{document}